# A STRUCTURAL THEORY OF EVERYTHING

Brian D. Josephson
Mind–Matter Unification Project,
Cavendish Laboratory, J J Thomson Avenue, Cambridge CB3 0HE.

ABSTRACT

In this paper it is argued that Barad's *Agential Realism*, an approach to quantum mechanics originating in the philosophy of Niels Bohr, can be the basis of a 'theory of everything' consistent with a proposal of Wheeler that observer-participancy is the foundation of everything. On the one hand, agential realism can be grounded in models of self-organisation such as the hypercycles of Eigen, while on the other agential realism, by virtue of the 'discursive practices' that constitute one aspect of the theory, implies the possibility of the generation of physical phenomena through acts of specification originating at a more fundamental level. Included in phenomena that may be generated by such a mechanism are the origin and evolution of life, and human capacities such as mathematical and musical intuition.

INTRODUCTION

The quest for a 'theory of everything' may be said to have begun with Newtonian mechanics, followed by Maxwell's electromagnetic theory, then basic quantum mechanics, later quantum electrodynamics, then ultimately the 'Standard Model', which provides a unified description of the strong, weak and electromagnetic forces, and fits the experimental data well. But the task of including gravitation in the 'theory of everything' has not turned out so well, the outcome being theories that are mathematically elegant, but where there is little or no contact with experiment. Other questions also arise in connection with the idea of a theory of everything, such as whether quantum mechanics as such is applicable to biology (1), and whether or not it provides an adequate account of observation.

The difficulties cited above may be of little interest to the 'working physicist', but they could have a similar status to that of the 'small clouds on the horizon' that beset classical physics at one time, ultimately leading to it being superseded by quantum theory. Those difficulties accordingly merit closer attention. The approach that seems most promising in this context is the realist point of view characterising the 'agential realism' of Karen Barad (2), and the 'reality of possibility' concept of Ruth Kastner (3). The realist position differs from the usual one adopted in quantum mechanics in that it attempts to describe what actually happens in the case of *individual events*, rather than simply computing averages. The difference is apparent in the case of a typical high-energy physics experiment in which large numbers of individual events are observed. Quantum theory addresses only statistical averages, whereas one can imagine instead a theory that can describe what happens individual events. In confining oneself to statistics as is the norm, one may be missing crucial information, as would indeed happen in sciences such as astronomy. This would clearly be the case in astronomy, where for example a statistical approach to meteor showers would ignore the occasional peaks.

 The present paper attempts to go beyond the formulations of Barad and Kastner in two ways. First of all, a close affinity will be demonstrated between agential realism and certain classical models of self-organisation, such as Eigen's hypercycle model, with the conclusion that the characteristic features of the quantum domain may merely involve some possibility within the classical domain, characterised by a particular kind of order. Quantum mechanics would still remain an important area of study, but would no longer be regarded as fundamental, and study of this specific type of order would then replace particle physics as currently practiced as the premier field of study in physics. Further, in this connection it will be shown how a particular aspect of agentive realism, namely the evolution of discursive practices, can account for the emergence of specific physical phenomena, in rather the same way that discourse within a technical community leads to technical advances on the human plane.  Such a picture would explain physical reality in a way totally different from the conventional one where some specific calculation would account for the details of phenomena.

INDETERMINACY

Barad's agential realism, an attempt to make sense of the quantum domain in realist terms, is a development of ideas due to Niels Bohr, where a central role is played by the concept of indeterminacy.  Bohr argued that details of the quantum domain are in general not merely *uncertain*, relating to mere ignorance of the details, but *indefinite* or *indeterminate*, in the sense that treating values of variables as definite would be contradictory.  For example, when a beam of light is split into two by a beam splitter and combined in such a way as to produce an observable interference pattern, we may be inclined, considering photons as particles, to ask which path was taken by an individual detected photon.  But determining which path a given photon had taken, using a suitably placed photon detector, would prevent interference.  This contradiction implies that the question of which path a particular photon took has no acceptable answer.  Bohr infers from this that one should not presume that particular assertions about a system of interest are necessarily meaningful; rather, what can be stated meaningfully is a function of the context.  Quantum particles are not like classical particles, where attributes can be presumed regardless of the context.

AGENTIAL REALISM

Barad considers it a defect of Bohr's philosophy as stated above that he gives too much priority to the situation of the physics experiment, thereby introducing an artificial distinction between experiments and the rest of nature.  Furthermore, he ignores the question of *what is the case* (ontology), considering only *what can be known* (epistemology). To address these deficiencies, Barad argues that what is real is the *phenomenon*, for example the action of measurement.  An aspect of such a phenomenon is the *agential cut*, referring to a separation following measurement between a measuring apparatus registering the outcome of a measurement (an *agent*), and something playing the role of an *object* of measurement.  This split is something that can happen in the context of a measurement, but the process described is hypothesised to be a more general one, happening under given conditions, producing in effect a subject-object split.

Furthermore, as noted by Bohr, we would be wrong to treat what is being measured when a measurement is taking place as something abstract such as position; rather the physical apparatus related to the abstraction is primary, and only as a limiting case does the measurement that the apparatus delivers correspond exactly to a concept that we may have such as position or momentum.  In the case of scientific measurements, an existing concept does guide the design of the measuring instrument, but this need not be the case in general; rather, a concept associated with an agent emerges over time as the agent evolves.

We can, if we wish, view the situation in the following terms.  A quantum particle is not like a classical particle, but more like a swarm of insects or birds, where while we can be definite that the entity exists, ascribing to the entity concerned it a definite position or velocity is not permissible.  Further, there can at times be a phenomenon analogous to a cut, which involves the appearance of a further distinguishable entity, or more generally structure of some complexity.  Such structures cannot be treated as autonomous entities, but rather as aspects of the total phenomenon and contingent upon it (cf. the situation in condensed matter physics, where it is understood that both the excitations (quasi-particles), and their interactions, are a function of the background medium). Dependence on the background applies equally to *relationships* between structures, which may also be characterised as phenomena; such relationships are primary in that they influence entities that are related; in Barad's words, *relationships precede relata*.

DISCURSIVE PRACTICE

The analysis can now be taken one step further, taking into account the fact that the activity of one structure may systematically influence others. This introduces the idea of a distinction between *information* and the *influence* of that information. Such a distinction involves a kind of symmetry breaking, where one entity starts to exert control over others, so it becomes a supplier of information to other systems. A further subtlety is Barad's concept of *discursive practice*. For information transfer to be able to be effective at achieving some outcome, information must be processed in accord with specific rules, and it is the specific behaviour associated with the application of such rules that Barad terms discursive practice, which has a degree of persistence not applicable to the acts of information exchange themselves.

We now consider discursive practice and its evolution in more detail. For communication between two entities to be effective, they must share a language or discursive system, an integrated entity that may involve and evolve many components, some of which involve discursive activity itself, whilst others relate this discursive activity to the context of its use, in other words to non-discursive practices. Discursive activity is a form of measurement process, with discourse relating to what control may be possible. Thus evolution of discursive practice is related to the emergence of new mechanisms for control.

Discursive practices are closely linked with *concepts*. Earlier we noted Bohr's comment that while position, for example, appears to be a well-defined concept, a word such as position has meaning only to the extent that we are able to say what the position of something is. Discourse adds precision to this state of affairs, as for example by the way it permits the formulation of rules. In the case of human society, discursive practices have reached a very high level of development, and our discussion of the emergence of physical laws will invoke the possibility of similar evolutionary development in the quantum domain.

THE ACTIVE ASPECT OF MATTER

Barad views matter as something active, configuring its surroundings so as to continue to function effectively ('concerning itself with what matters'). Activity as such is by no means an unorthodox presumption, since omnipresent activity is familiar both in classical physics, in the context of molecular motion at finite temperatures, and in the case of quantum-mechanical zero-point motion. And in the biological context, matter configuring its surroundings appropriately is equally familiar. What is new here is simply the idea that the quantum domain involves similar configurability.

Eigen's discussions of biological self-organisation (4) are relevant in this context. The question he addresses is how the complicated molecular systems found in biology can sustain themselves and also evolve. He concludes that a structure known as a hypercycle plays a crucial role, being able for example to utilise small selective advantages and to be able to evolve quickly on the basis of these selective advantages. Such a system, involving a closed cycle of information carriers coding for functional systems that generate the information carrier next in the cycle, can also develop branches exhibiting various functionalities, with a common element allowing the branches to coexist. Computer simulations have demonstrated the validity of such conclusions.

Eigen's models involve molecules, but the principles involved could be equally relevant in the context of agential realism, provided that structures with corresponding features exist. There seem to be no fundamental reasons why this should not be the case, especially in a system close to an instability, where non-linearity would be conducive to structure formation.

Hankey (5) has arrived at conclusions somewhat similar to those of Eigen, along the following lines. A system that performs a sequence of changes can develop an instability if some end product is similar to the starting situation, thereby allowing the cycle to be repeated. The existence of such cycles can be probed by a general increase in the effective gain, and when such cycles emerge changes can be made to enhance the extent to which they are self-sustaining. Such self-sustaining cycles are then readily available for use in other situations. Learning in general is known to involve similar cyclic processes (i.e. the repetition that typically forms an integral part of the learning process).

RELATIONSHIP TO THE POSSIBILIST INTERPRETATION OF QUANTUM MECHANICS

In Kastner's 'possibilist interpretation' of quantum mechanics (3), the collapse of the wave function associated with observation is interpreted as the outcome of a *transaction* with a *possibility*, which outcome makes something definite, in the same way as in the present analysis. Possibilities are things that can be considered real, but are not necessarily *actualized*, which process Kastner equates with existing 'within the observable spacetime theater'. The theory developed here can accommodate possibilities but in a different way, namely as *aspects of discourse* rather than something real. In this way, Kastner's transactions with possibilities translate into discourse involving possibilities, of which some may subsequently be realised.

EVOLUTIONARY AGENTIAL REALISM

John Archibald Wheeler, in his article 'Law without Law' (6), explored the hypothesis that the 'observer-participancy' of quantum mechanics (equivalent to the measurement process discussed in the above) could be the basis of a new science that could replace the existing 'imposing structure of science'. The above considerations leave us in a position to understand how this might be achieved. The new picture involves agents evolving more and more advanced concepts over time, instantiated by evolving discursive practices along the same lines as those involved with cultural evolution. The basic mechanics of evolution would be the same, only the context in which that evolution would take place being different. With cultural evolution, we start with regular physics and chemistry, and with nervous systems already preprogrammed with mechanisms for developing along particular lines. Here we start with a system that knows essentially nothing, but which can explore and learn, provided the initial physics supports the kinds of structures and relationships that are required.

As to details, what we know about how children come to make sense of the world is arguably relevant, since they also start knowing almost nothing. Piaget (7) made systematic studies of the incremental changes that occur during development, each of which can be identified with the coming into operation, and utilisation, of a new module. An example is the development of the object concept: at a certain stage in a child's development an object that is covered up so it is no longer visible is treated as if it no longer exists: for example it will stop trying to reach it if it had been trying to reach it previously, but later new resources become available and actions directed towards the object continue even when it cannot be seen. There is a clear logic (8) to the sequence of the incremental changes, as a consequence of the way existing skills need to be operational before new ones can be acquired. The computer model of Osborne (9) exemplifies such an incremental process.

As regards the actual mechanics, what has been discussed so far suggests the following picture. From an initial hypercycle the developments envisaged by Eigen and Hankey can occur, with new concepts being established through the mechanisms envisaged by Bohr and Barad, whereby some apparatus becomes more and more representative of particular concepts over time. The fact that this process involves the use of the mechanisms corresponding to a concept implies in effect that it is the most useful concepts that get developed in this way. Familiar examples in childhood development are those of an object, and number, which get clearly defined though apparatus that deals with reality in such terms.

As regards space, we recall that geometry developed originally through the study of manipulations involving actual movements in space, and we can imagine the agents of agential realism acquiring spatial concepts in the same way. Barad considers space not as 'a collection of preexisting points set out in a fixed geometry, a container … for matter to inhabit', but rather as the concomitant of a collection of interactions within a system (referred to by her as *intra*-actions), defined as a topology. How this topological structure is related to ordinary geometry is not discussed, but given the relevance that symmetry groups have in geometry, one possibility would be that the concepts of transformation and invariance play a role, combined with group properties defined in terms of transformations, which together can serve to delineate a geometry. Our subtle agents are presumed to be able to work on the basis of such concepts to shape a geometry to fit particular specifications.

In the end, discourse can come to cover the question of techniques for creating particular kinds of universe, and particular kinds of life in such universes. Such partly directed behaviour could have relevance to questions such as the origin and evolution of life where, perhaps as a consequence of the influence of Monod's writings (10), it has traditionally been assumed that there is no preferred direction or meaning in the universe.

CONCLUSIONS

The picture we have arrived at, through the conjunction of a number of theoretical analyses, is that of a quantum reality containing entities that can come, over time, in effect to understand their reality, and to manipulate it, in the same way that a child learns to do this as it grows up.

The physical support for these processes is the equivalent of the evolving self-sustaining cycles discussed by Eigen and Hankey, which picture corresponds closely to that derived by Barad starting from Bohr's analyses of quantum reality. The conclusion is that under certain conditions matter can organise itself in ways according with agential realism, while at the same time agential realism, through the evolution of discursive practices, implies the possibility of primordial matter being able to create universes with specified physical laws, and being able to influence phenomena in these universes.

Support for this picture may come from the existence of human skills such as mathematical and musical intuition, which point to the ability to access levels of understanding that are hard to explain on conventional grounds (11). There are some parallels between the present situation and that which prevailed at the time when the atomic theory was being developed, in that the latter postulated a deeper reality capable of accounting, in simple terms, for a number of phenomena, such as the specific heat and thermal conductivity of gases that were, at that time, characterised in purely mathematical terms, with no underlying model. Taking the hypothesised deeper reality seriously led to considerable advances in science.

While this the view presented here is very different to the usual one, there is nothing irrational about it; it merely contradicts ideas that are commonly held, and these ideas would be irrational themselves if held as dogma. It must be for the future to examine the ideas presented here in detail, and to see how well they hold up to examination.

ACKNOWLEDGEMENTS

I wish to thank Ilexa Yardley for discussions of her *Circular Theory*, which has ideas closely related to the approach adopted here, though presented in different terms, and Alex Hankey for discussion of ideas also related to the present ones. I am indebted also to Plamen Simeonov, Madan Thangavelu and Steven M. Rosen for stimulating discussions.